# Segmentation of Infrared Breast Images Using MultiResUnet Neural Networks

Ange Lou[1], Shuyue Guan[2], Nada Kamona[2], Murray Loew[2]
Medical Imaging and Image Analysis Laboratory
[1]Department of Electrical and Computer Engineering, [2]Department of Biomedical Engineering
The George Washington University Medical Center
Washington DC, USA
{angelou, frankshuyueguan, nkamona, loew}@gwu.edu

*Abstract* — Breast cancer is the second leading cause of death for women in the U.S. Early detection of breast cancer is key to higher survival rates of breast cancer patients. We are investigating infrared (IR) thermography as a noninvasive adjunct to mammography for breast cancer screening. IR imaging is radiation-free, pain-free, and non-contact. Automatic segmentation of the breast area from the acquired full-size breast IR images will help limit the area for tumor search, as well as reduce the time and effort costs of manual segmentation. Autoencoder-like convolutional and deconvolutional neural networks (C-DCNN) had been applied to automatically segment the breast area in IR images in previous studies. In this study, we applied a state-of-the-art deep-learning segmentation model, MultiResUnet, which consists of an encoder part to capture features and a decoder part for precise localization. It was used to segment the breast area by using a set of breast IR images, collected in our pilot study by imaging breast cancer patients and normal volunteers with a thermal infrared camera (N2 Imager). The database we used has 450 images, acquired from 14 patients and 16 volunteers. We used a thresholding method to remove interference in the raw images and remapped them from the original 16-bit to 8-bit, and then cropped and segmented the 8-bit images manually. Experiments using leave-one-out cross-validation (LOOCV) and comparison with the ground-truth images by using Tanimoto similarity show that the average accuracy of MultiResUnet is 91.47%, which is about 2% higher than that of the autoencoder. MultiResUnet offers a better approach to segment breast IR images than our previous model.

*Keywords — infrared breast segmentation; MultiResUnet; deep-learning; autoencoder; computer-aided diagnosis.*

## I. INTRODUCTION

Breast cancer is estimated to account for more than 40,000 deaths of women [1]. Early detection is key to higher survival rates for breast cancer patients [2]. Mammography has been the gold standard for breast cancer screening [3], and despite its success in reducing mortality rates, its performance is influenced by breast density. Moreover, mammography increases radiation risk and is less effective for younger women [4] [5]. On the other hand, thermography is non-radiation, non-contact and low-cost, making it a promising noninvasive adjunct to mammography for breast cancer screening [5]. Studies show that modern analysis techniques based on thermography have the potential to detect breast cancer reliably [6].

We are conducting a study in which we are imaging patients diagnosed with breast cancer using a thermal infrared camera. As illustrated in Figure 1, thermal images may contain a noisy background and other unnecessary areas like the neck, shoulders and lower abdomen, which may ultimately influence automatic tumor search algorithms. Accurate search for tumor regions requires that the background be removed from the infrared images, and the breast region should then be segmented from the reduced of image [7]. This will minimize the region of interest used for tumor search. Finding accurate boundaries of the breast region is challenging due to the inherent limitations of thermal images (including the low contrast, low signal-to-noise ratio [8], and lack of clear edges), as well as great variations in breast shapes and sizes. Previous approaches proposed for breast region segmentation in thermal images included thresholding methods, region-based methods, Snake algorithms [7] and level sets [9].

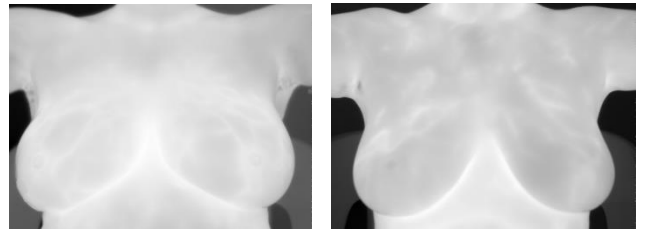

Figure 1. Full thermal raw images of two patients, including the neck, shoulder, abdomen, background and chair.

Recently, deep learning has become a popular method for image segmentation. The U-Net, which is the segmentation model derived from convolutional and deconvolutional neural networks (C-DCNNs), is a powerful medical image segmentation model. For instance, the U-Net has been widely used in retinal vessels and serial section Transmission Electron Microscopy (ssTEM) images segmentation [10]. In this study, we applied a more powerful state-of-the-art deep learning segmentation model: MultiResUnet, an extension of the U-Net, to compare with our previous work.

In the following experiments, we used leave-one-out cross-validation and compared with ground-truth images using Tanimoto similarity. The results show the average accuracies of C-DCNN and MultiResUnet to be 89.7% and 91.5%, respectively. The results demonstrate the potential of MultiResUnet as a better approach to segment infrared breast images than our previous model.

## II. MATERIALS

### A. Breast Thermography

Tumor growth is accompanied by increased growth of blood vessels, known as angiogenesis. Blood vessels increase oxygen and nutrient delivery to the growing tumor, and this increased blood flow cause a rise in local temperature of that region compared to the temperature of the surrounding tissues. Thermography has the potential to detect those elevated skin temperatures that arise from the increased blood flow.

We collected infrared images using the N2 Imager (N2 Imaging System, Irvine, Calif.). The camera detects wavelengths in the long infrared region in the electromagnetic spectrum (8-12 microns); humans at normal body temperature radiate thermal infrared at wavelengths around 10 microns. The camera has a 640 by 480 array of 17-micrometer pixels, and a stated thermal resolution of 18.6 mK.

Pilot studies are currently being conducted in collaboration with the Breast Clinic at the GW Medical Faculty Associates. Patients diagnosed with breast cancer are imaged with the infrared camera for a total time of 15 minutes to observe cool down of the breast tissue. This dynamic thermography monitors the temporal behavior of breast thermal patterns, which in our case is the cool-down of breast tissue over time. The patient sits still with both arms raised on two arm supports, with the camera positioned approximately 25 inches away from the patient (frontal view). Imaging starts immediately after the patient undresses, capturing images every minute as the patient's skin cools down. Figure 2 shows the environment for our breast image acquisition.

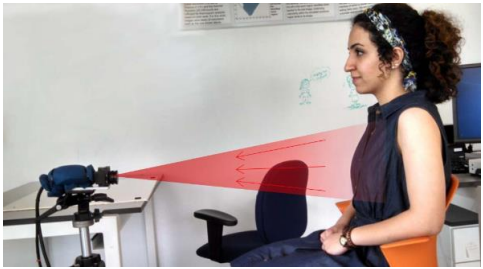

Figure 2. Our breast infrared thermography system.

### B. Image Collection and Pre-processing

Our database contains 450 infrared images from 14 patients and 16 normal healthy volunteers; all of them contain background objects and noise. Each participant was imaged for a total time of 15 minutes, capturing one image every minute (15 images per participant total). The 16-bit images were converted to 8-bit, and the interferences were eliminated at the same time. We applied an averaging filter with kernel size 101 by 101 to smooth the raw images and used the Otsu [16] algorithm to find proper thresholds. Those thresholds are modified by adding a compensation value. We removed the interferences by using this new threshold and remapped the raw images to 8-bit. The preprocessed 8-bit images are shown in Figure 3(b). Compared with raw images, the new preprocessed images have cleaner backgrounds and better contrast.

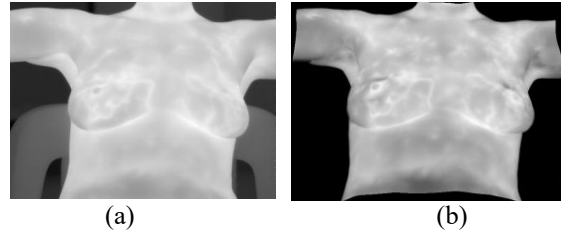

Figure 3. Removing background of the raw IR images: (a) raw image of patient 3 with background and noise, (b) new preprocessed image after removing background.

We cropped the newly preprocessed 8-bit images manually to remove the upper and lower regions (neck and abdomen). Then, we segmented them manually by tracing the breast curvature and extracted the breast region to generate the ground-truth segmentation (Figure 4).

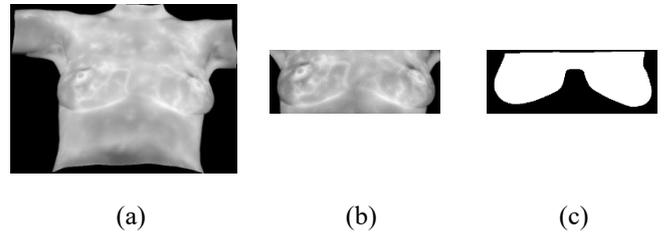

Figure 4. Preprocessing of IR images: (a) 8-bit IR image, (b) manual rectangular crop to remove shoulders and abdomen, (c) ground-truth manual segmented image.

## III. METHODS

### A. Segmentation Architecture

The segmentation model C-DCNN [13] we had used in a previous study connects a CNN and DCNN. It consists of six convolutional layers, two max-pooling, two up-sampling layers, three fully connected layers, and one flattening layer. Convolutional parts transform a 2-D image to a feature vector and deconvolutional parts then convert a vector to an image. Overall, the model transforms images to images. In this study, we used MultiResUnet, which is based on the classical U-Net architecture. The U-Net model also contains two parts: encoder and decoder [11]. The encoder follows the typical architecture of a convolution network, and the decoder is the deconvolution network. In the decoder, we combined the feature map which is cropped correspondingly from the encoder. The MultiResUnet (Figure 5) consists of two parts: MultiRes Block and Res-Path [12].

In the MultiRes Block, we incorporated $5 \times 5$ and $7 \times 7$ convolution operations in parallel with the $3 \times 3$ convolution (Figure 6 (a)). We took the output from these three convolution blocks and concatenated them to extract the spatial features from different scales. However, computing $5 \times 5$ and $7 \times 7$ convolution is computationally expensive.

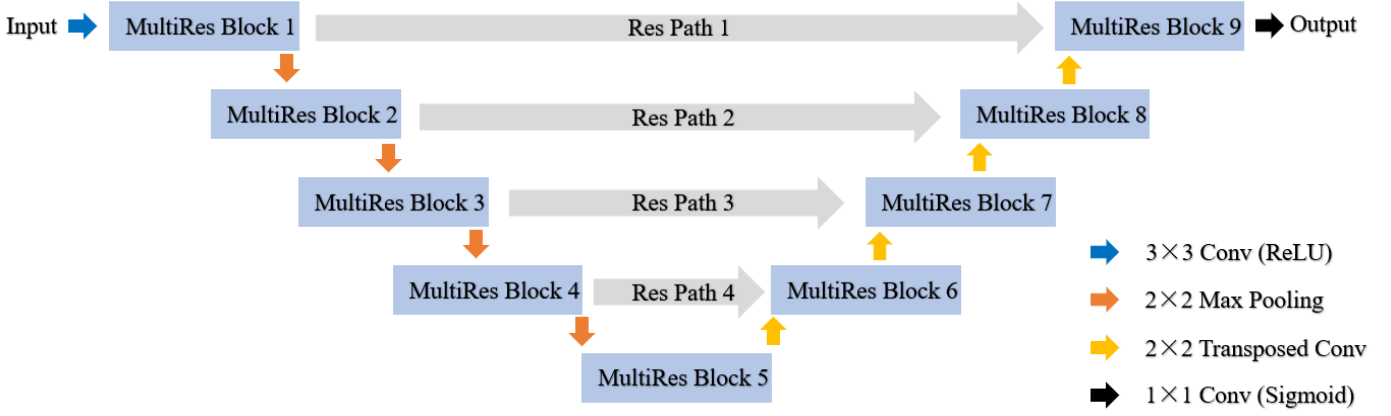

Figure 5. Architecture of MultiResUnet.

To reduce the computation, we applied a chain of 3 × 3 convolution kernel to replace the 5 × 5 and 7 × 7 kernels, which can be simplified as shown in Figure 6 (b). We also added a residual connection and applied 1 × 1 convolution layers to provide additional spatial information. We combined those structures to build MultiResUnet (Figure 6 (c)).

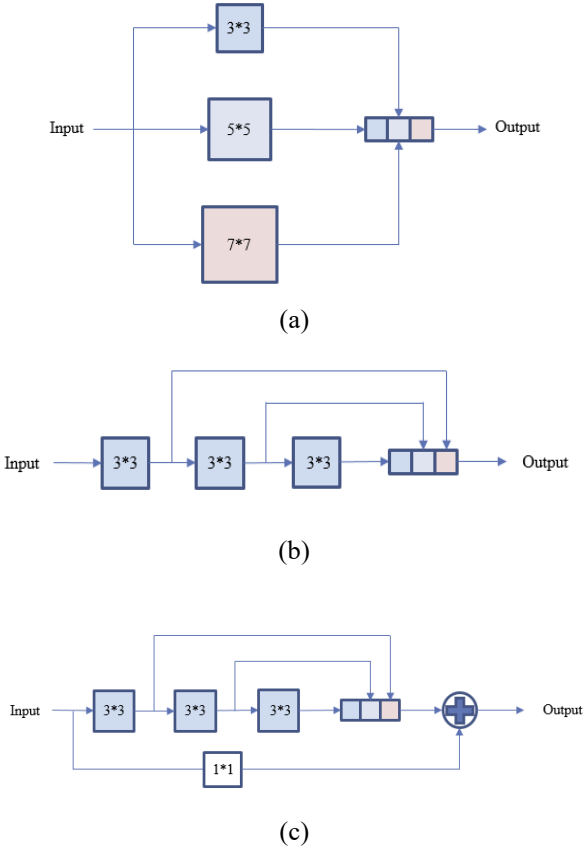

Figure 6. MultiRes Block: (a) inception model, (b) simplified inception model, (c) add residual connection to build MultiRes Block.

The Res-Path (Figure 7), instead of simply concatenating the feature maps from the encoder part to decoder part, passes them through a chain of 3 × 3 convolution layers with residual connections, which then are concatenated with the decoder feature.

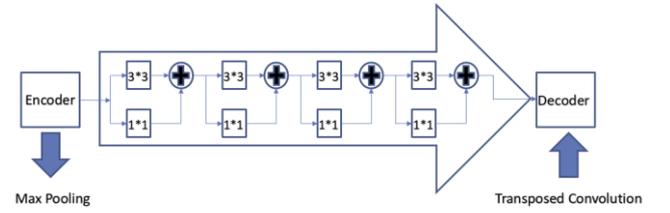

Figure 7. Res-Path.

*B. Experiment*

To test the performance of MultiResUnet, we applied the leave-one-case-out experiment. We withheld 15 images of one subject from the database as the testing set, and all other 29 subjects with a total of 435 images as the training set. This experiment approach shows the performances of various models to segment a brand-new breast case that was not used during training. Since the final segmentation output is not binary, we chose the Tanimoto similarity as the metric for comparing the models' segmentation results to the manual ground truth segmentation. This helps avoid thresholding errors introduced by other metrics such as the Intersection over Union (IoU).

Our neural network model was on the Keras API backend on TensorFlow [14], with the development environment for Python being Anaconda 3. We set 30 epochs for training with batch-size 3 and chose binary cross entropy as the loss function. The optimizer is Adam [15] using default parameters with learning rate 1e-3. Training costs 17 seconds per epoch by using a single Nvidia RTX 2070 GPU.

## IV. RESULTS AND EVALUATION

The average accuracies of C-DCNN, U-Net, and MultiResUnet are 89.7%, 89.8%, and 91.5%, respectively, after applying leave-one-out experiments for the three models. As shown in Figure 8, the MultiResUnet model performs better than the other models for most test cases. Table 1 shows the average accuracies and time cost of the three models.

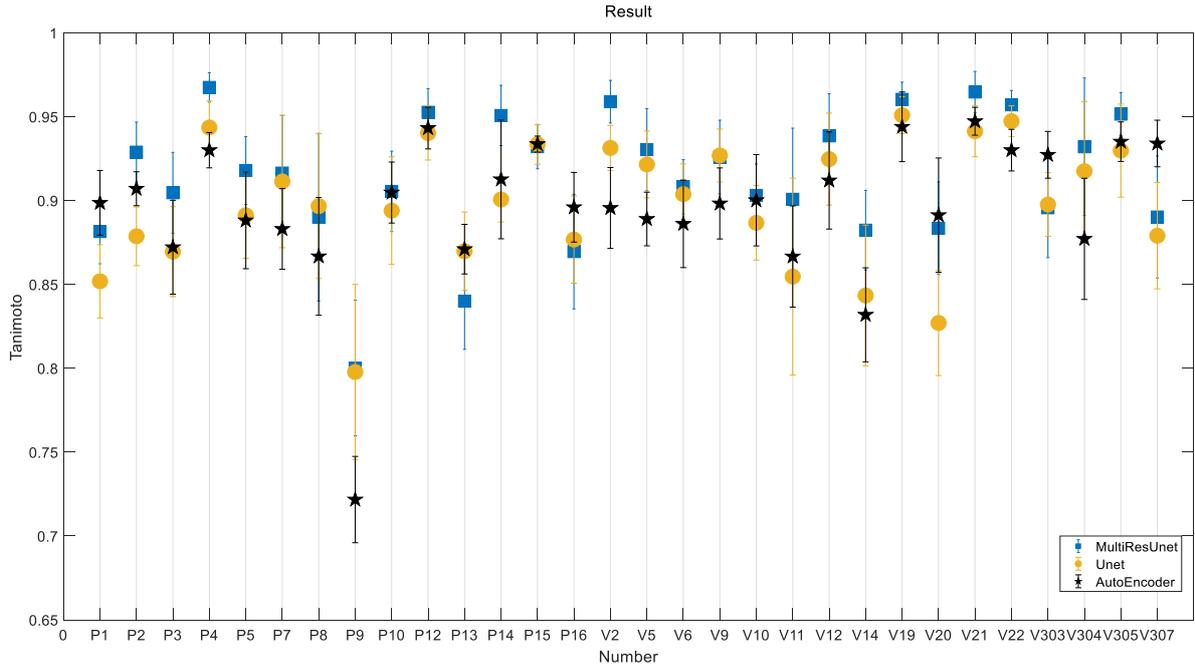

Figure 8. Breast segmentation results of the models for each subject, using the Tanimoto metric for accuracy. On the x-axis, the "P" and "V" corresponds to patient and volunteer.

TABLE 1. AVERAGE SEGMENTATION ACCURACY FOR 30 SAMPLES

| Model | C-DCNN | U-Net | MultiResUnet |
|---|---|---|---|
| Accuracy (Tanimoto) | 89.7% | 89.8% | 91.5% |
| Seconds per epoch | 5 | 15 | 17 |

Although the Tanimoto results of C-DCNN and U-Net are similar, U-Net provides a more accurate breast boundary as shown in Figure 8. For volunteer 7 shown in Figure 9, the average (over the 15-minute imaging period) accuracy of C-DCNN, U-Net, and MultiResUnet are 91.20%, 92.47%, and 93.86%, respectively, all being high in value. MultiResUnet gives the best segmentation results compared to the other models, however, because it combines features of various resolutions.

The errors in the segmentation output are due to variations among subjects and the errors in the manual segmentation – particularly in the upper boundary of the ground-truth image. Errors in the ground-truth segmentation include inconsistency of drawing the upper and middle boundaries of the breast region. For volunteer 7, all models show the upper boundary is a line close to the upper boundary of cropped images; the line, however, is lower in the ground-truth segmentation, resulting in a decrease in the accuracy of three models compared to other cases, despite the fact that this is still conceived as a good segmentation output that captures the breast region of interest.

Moreover, breast size influences the segmentation accuracy. For volunteer 14 shown in Figure 10, the average accuracies of C-DCNN, U-Net and MultiResUnet are 83.19%, 84.34%, and 88.24%, respectively; this lower performance is most likely because the size of the breast is substantially smaller than the other cases used for training, which poses a problem for a simple architecture model. This will be resolved in the future by increasing the size and diversity of the training set to have a greater representation of breast shapes and sizes.

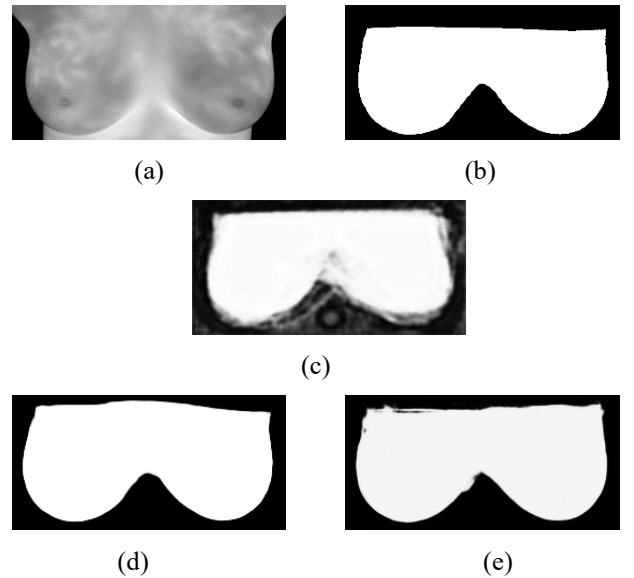

Figure 9. Segmentation results of volunteer 7. (a) Cropped Image (b) Manual ground-truth (c) C-DCNN (d) U-Net (e) MultiResUnet

V. DISCUSSION

A. Comparison of the Two Models

The 30 subjects' average Tanimoto accuracies of C-DCNN, U-Net and MultiResUnet are 89.7%, 89.8%, and 91.5%, respectively. The results illustrate that a more powerful and complex model can extract richer features to improve segmentation of the breast region. The results for volunteer 7 show that U-Net series models provide clearer and better results: U-Net series models recover better boundary and corner details.

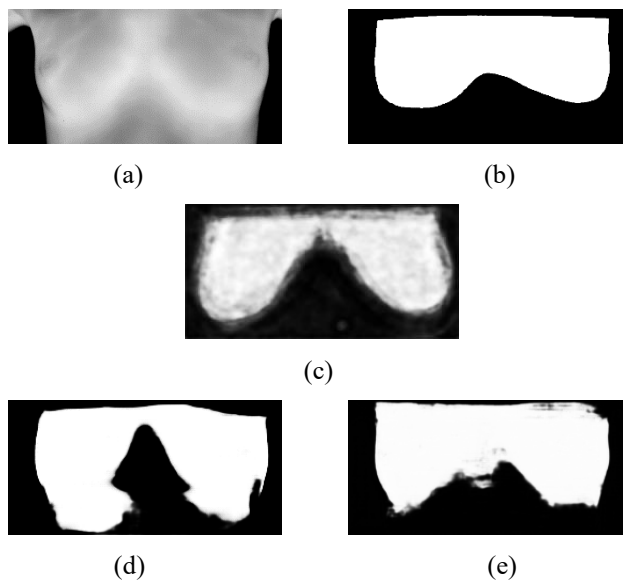

Figure 10. Segmentation results of volunteer 14. (a) Cropped Image (b) Manual ground-truth (c) C-DCNN (d) U-Net (e) MultiResUnet.

*B. Improvements and Feature Work*

A larger database was used in this experiment compared with our previous study. We applied U-Net based model to compare with the C-DCNN in order to solve the small-breast problems. The results show that MultiResUnet provides a more accurate boundary and is more stable to train. We also used a new method, Tanimoto-similarity, instead of IoU to measure the segmentation accuracy. This measurement avoids additional errors from the thresholding process of gray-level result.

Results show that U-Net series models can capture essential features of breast regions and delineate them in the testing dataset through cross-validation and comparison with the ground-truth images. Adding residual connections to make the model deeper, and applying multi-resolution features (MultiResUnet) improves the average accuracy from 89.7% to 91.47% based on Tanimoto-similarity. A deeper and richer-feature model shows better segmentation performance. Reasonable preprocessing and a greater number of training data could improve future performance of deep-learning models. For the same subject, a MultiResUnet model can output more stable segmentation than manual segmentation.

In future work, one approach to improve the segmentation accuracy is to combine a deep-learning model with other preprocessing methods, such as data augmentation, to enlarge the dataset and avoid overfitting. In the MultiResUnet, we can apply more powerful blocks as encoder and decoder to extract effective features, such as the blocks in GoogleNet [18], Inception-v4 [19], and ResNet [20].

VI. CONCLUSION

Segmenting and isolating the breast region in thermal images is a necessary pre-processing step for automated computer algorithms to detect breast cancer tumor regions. Manual segmentation is costly and is likely to have inter- and intra-observer errors. On the other hand, automatic segmentation using stable models has the potential to provide accurate and repeatable breast segmentation with less time and effort.